\documentclass[runningheads]{llncs}
\usepackage{url}
\usepackage{graphicx}
\usepackage{amsmath}
\usepackage{amssymb}
\usepackage{booktabs}
\usepackage{float}
\usepackage{multirow}
\usepackage{tikz}
\usepackage{bbding}
\usetikzlibrary{shapes.geometric, arrows}
\usepackage{xcolor}
\newcommand{\Defender}[1]{\textcolor{blue}{\textit{#1}}}
\usepackage{algorithm}
\usepackage[noend]{algpseudocode}
\floatname{algorithm}{Procedure}
 
\begin{document}
\raggedbottom

\renewcommand{\topfraction}{0.9}
\renewcommand{\bottomfraction}{0.8}
\renewcommand{\textfraction}{0.07}
\renewcommand{\floatpagefraction}{0.85}

\title{Covert System Identification Under Timing Constraints:
A Static Linear Model and Its Detectability Implications}

\author{
Victor Mattos\inst{1}\,\Envelope
\and Guilherme Sobreira\inst{1}
\and Amit Bhaya\inst{1}
\and Alan Oliveira de Sá\inst{5}
\and Charles B. Prado\inst{4}
\and Raphael Machado\inst{2, 4}
\and Gaurav Srivastava\inst{3}
\and Daniel~S.~Menasche\inst{1}
}

\authorrunning{V. Mattos et al.}

\titlerunning{Covert System Identification Under Timing Constraints}

\institute{
Universidade Federal do Rio de Janeiro (UFRJ), Rio de Janeiro, Brazil\\\email{victor.mattos@coppe.ufrj.br}
\and
Universidade Federal Fluminense (UFF), Niterói, Brazil
\and
Siemens, Princeton, NJ, USA
\and
Instituto Nacional de Metrologia, Qualidade e Tecnologia (Inmetro), Brazil
\and
LASIGE, Faculdade de Ciências, Universidade de Lisboa, Lisbon, Portugal
}

\maketitle

\begin{abstract}
This paper studies covert system identification in cyber-physical systems, where an attacker infers plant dynamics while attempting to remain stealthy. We introduce a lightweight static linear model in which the time to execute a setpoint change is proportional to the commanded speed difference. On a Siemens PROFINET testbed this gain captures the drive's ramp response accurately but not its terminal overshoot. That omitted transient becomes the attacker's main exposure: on real traces, the residual left by the linear masking procedure is detected by two anomaly monitors, Process-Aware Stealthy-Attack Detection (PASAD) and the CUSUM change detector. Across the cascade identification and staircase attack strategies defined in the paper, the stealth boundary relates to the plant's own settling transients, a structural limit for this class of models under the adopted detector-calibration regime. Comparison against dynamic models obtained by N4SID, trained at multiple excitation amplitudes, shows that the evaluated N4SID models do not improve stealth over the structure-informed ramp model. Since per-cycle computation is negligible for both procedures, the binding trade-off is not inference cost but whether the attacker can acquire and refresh a model without injecting detectable perturbations.
\end{abstract}



\section{Introduction}
\label{sec:intro}

System identification is a fundamental tool in cyber-physical systems, enabling the estimation of plant dynamics from observed input-output data. Although traditionally used for control and monitoring, it can also serve an attacker who infers system behavior to support stealthy manipulation~\cite{alanCovertAttacksCyberPhysical2017}.

In this paper, we study \emph{covert system identification}, that is, system identification performed under strict stealth constraints. We consider an attacker deployed as a man-in-the-middle (MitM) bridge between a controller and an actuator. The attacker's objective is two-phase: first acquire and refine a model of the plant under stealth constraints, then use that model to mask a covert manipulation, such as an unauthorized change to the motor's commanded speed, while attempting to remain undetected.

This setting poses an apparent trade-off. A lightweight model is computationally cheap to deploy in real-time but may be too simple to reproduce the plant's behavior, leaking modeling error to an anomaly detector. A more complex model would capture more of the real behavior but is more expensive to compute, and computation inside a real-time loop itself can be observable. We begin with the lightweight end: a simple static linear gain model relating setpoint variation to execution time, computationally efficient and empirically accurate through the commanded ramp, but unable to capture the overshoot. 

\medskip

This paper makes the following contributions:

\begin{itemize}
    \item We study covert system identification under timing constraints and introduce a lightweight static linear gain model for attacker-side inference, accurate through the commanded ramp ($R^2 \approx 0.999$) on a Siemens PROFINET testbed, deviating only at the plant's transient overshoot.
    \item We introduce two covert strategies: \emph{cascade} identification, an incremental masked model-refinement scheme, and the \emph{staircase} attack, a large covert speed change decomposed into small masked increments. Across both, the masked residual exposes the attacker to two anomaly detectors --- Process-Aware Stealthy-Attack Detection (PASAD) and CUSUM.
    \item We compare the static linear gain model against dynamic linear models obtained by N4SID, trained at seven active-identification amplitudes, and show that none of the evaluated N4SID masks extends the static map's stealth boundary in the attack phase.
    \item We quantify the computational side of the trade-off and show that per-cycle inference cost is negligible for both masks; the relevant difference between the static and dynamic models lies in identification cost and in the active perturbations needed to keep a model current.
\end{itemize}

\medskip

The remainder of the paper is organized as follows. Section~\ref{sec:related} reports related work.
Section~\ref{sec:arch} describes the system architecture and experimental setup. Section~\ref{sec:model} introduces the linear model and evaluates its accuracy. Section~\ref{sec:artifats} covers detectability at the control layer: it defines the overshoot residual, introduces two detectors, and evaluates both the identification phase (cascade strategy, Section~\ref{sec:cascade_detection}) and the attack phase (staircase strategy and N4SID comparison). Section~\ref{sec:discussion} connects the detection results to the identification-cost trade-off, and Section~\ref{sec:conclusion} concludes and outlines future work.

\section{Related Work}
\label{sec:related}

Our work sits at the intersection of two lines of work: on the attacker's side, covert attacks on cyber-physical systems and system identification under operational constraints; on the defender's side, intrusion detection. We review them in turn: first, attacker-side model acquisition and covert feedback manipulation; second, defender-side process monitoring and the role of model complexity in determining detectability.

\subsection{Covert and System Identification Attacks} 

Smith~\cite{smith} introduced \emph{covert misappropriation}, a covert attack in which an agent inserted in the feedback loop uses a model of the system to cancel its own effect on the sensor signal, so that the controller observes a fabricated yet consistent signal that hides an otherwise visible manipulation. We adapt this construction.
De Sá et al.~\cite{alanCovertAttacksCyberPhysical2017} propose a taxonomy of covert attacks on networked control systems and instantiate it with a two-stage attack: a metaheuristic system-identification stage that estimates the plant and controller transfer functions from observed loop signals, followed by a MitM stage that uses the identified model to degrade the process covertly. Our own prior work~\cite{mattos2025design} designed and detected covert MitM attacks on a water-treatment testbed. The present paper specializes that line of work to the system-identification phase itself, asking not only whether a covert feedback model can evade detection once deployed, but whether \emph{acquiring} that model under PROFINET's strict timing constraints is itself detectable, and at what complexity-versus-stealth trade-off.

When the process is \emph{already} excited by routine setpoint changes, the attacker can identify it passively and remain covert ~\cite{desaControllerDesignMitigation2018}. Our hardware study in this paper addresses the more difficult steady-state regime, where identification must be actively excited, and thus subject to possible detection. 

\subsection{Detection and Model Complexity}

On the defensive side, detectors that monitor the physical process, rather than network traffic alone, have proven effective against attacks that are stealthy at the network layer. PASAD~\cite{pasad} is the representative case used here: it is process-aware, it builds a subspace model of the plant's normal dynamics and scores each new observation window by its departure from that subspace. This posture responds to a line of work on stealthy attacks in networked control systems. Teixeira et al.~\cite{teixeira2015secure} characterize when a model-aware adversary can stay undetected, and what a defender must monitor to reveal it; false-data-injection attacks~\cite{liu2011false} formalize the null-space stealth condition that motivates process-side monitoring; and risk-and-detection frameworks for process control~\cite{cardenas2011attacks} situate such detectors operationally.

The attacker is evaluated against PASAD and, as a baseline, a classical CUSUM change detector~\cite{Montgomery_2020} applied to the same residual (Section~\ref{sec:artifats}). The dynamic alternative to our static model is a classical system-identification~\cite{ljung1999system} tool: N4SID~\cite{vanoverschee1994n4sid} recovers state-space dynamics at a substantially greater identification cost. This spectrum, from a static linear map to a fitted state-space model, is exactly the complexity-versus-cost space in which the trade-off is examined. Our contribution connects an attacker-side modeling decision---linear and memoryless versus higher-order dynamic---to its detectability under monitoring and to the cost of maintaining a sufficiently accurate model under real-time constraints.

\section{System Overview and Experimental Setup}
\label{sec:arch}

\begin{figure}[tbp]
    \centering
    \includegraphics[width=0.8\linewidth]{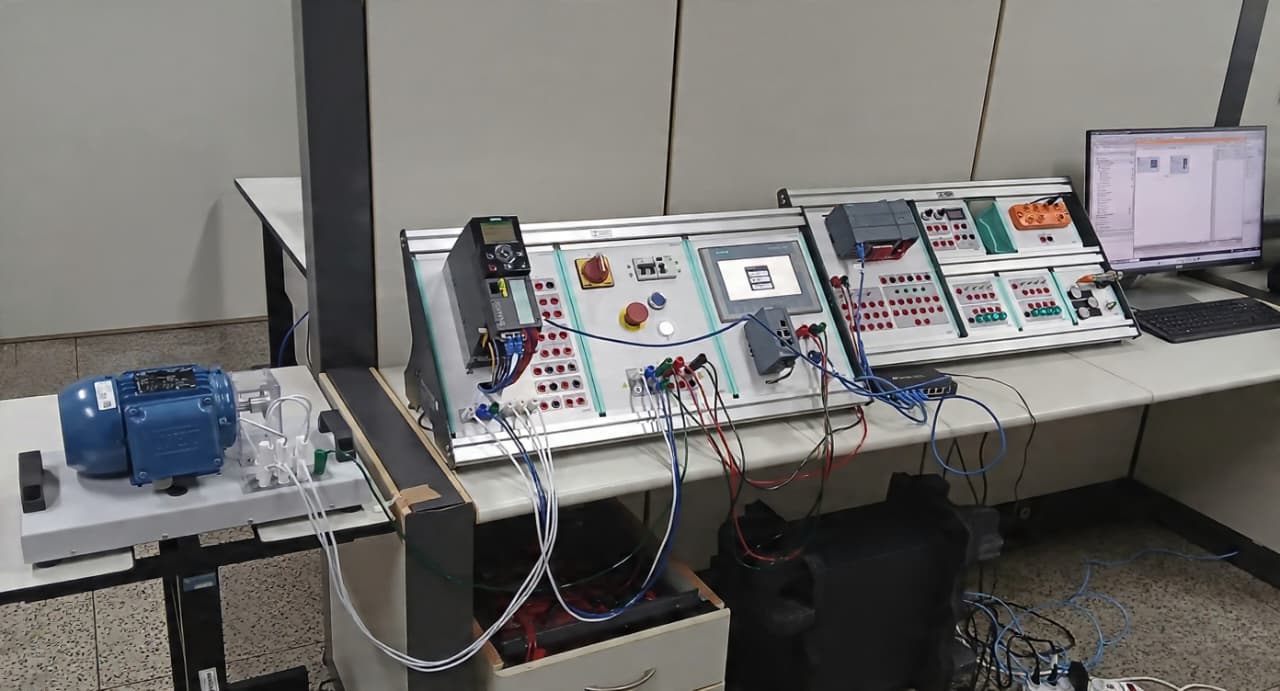}
    \caption{The experimental testbed. From left to right: induction motor; the didactic panel carrying the VFD and an operator HMI touch panel; the panel carrying the PLC; and the switch, which interconnects the PLC and the VFD. The workstation at right provides supervisory access. No MitM is present in this baseline configuration.}
    \label{fig:testbed}
\end{figure}

The experimental educational platform, shown in Fig.~\ref{fig:testbed}, consists of a control system for an industrial induction motor. The main components are:

\begin{enumerate} 

    \item \textbf{Programmable Logic Controller (PLC):} The PLC (Siemens S7-1214C) communicates with the Variable Frequency Drive (VFD) via the PROFINET industrial Ethernet protocol. In this experimental topology, the PLC acts as a supervisory controller: it transmits speed setpoints to the VFD and relays the drive's feedback to the operator without acting on the actual-speed word, so the covertness results below assume no PLC interlock is keyed on the motor's actual speed---no control or safety logic that would act on the very word the attacker rewrites~\cite[Chapter~3]{siemens_g120_fieldbus_2018}. The drive is responsible for the internal closed-loop regulation. Additionally, acceleration and deceleration ramp profiles are configured and executed locally within the VFD.
        
    \item \textbf{Variable Frequency Drive (VFD) --- the Inverter:} The VFD (Siemens SINAMICS G120) is configured in Speed Control mode utilizing Sensorless Vector Control (SLVC).

    \item \textbf{Induction Motor:} WEG W22, three-phase, rated power 0.55 kW, rated voltage 220~V, and nominal current 2.77~A.

    \item \textbf{Network Infrastructure:} TP-Link SG108E Easy Smart Switch, used to interconnect the PLC, VFD, and the attacker's station.

    \item \textbf{Attacker Station:} A workstation running Ubuntu 24.04.3 LTS on an Intel Core i5-12450H with 16~GB of RAM, using two physical Ethernet ports, one facing the PLC and one facing the switch. The same machine performs the packet interception and, in Section~\ref{sec:discussion}, the cost measurements.
\end{enumerate}

\subsection{PROFINET Process Data: Standard Telegram~1}
\label{sec:telegram}
The PLC and the VFD exchange data cyclically at a 2-ms rate, through the PROFIdrive profile using \emph{Standard Telegram~1}, defined in the drive function manual~\cite[Chapter~3]{siemens_g120_fieldbus_2018}. Table~\ref{tab:telegram1} lists the two 16-bit words (PZD-2/2) each cycle carries in each direction. The attacker considered in this work observes and rewrites exactly these four words.

 \begin{table}[htbp]
\centering
\caption{Standard Telegram~1 (PZD-2/2): four words exchanged each cycle.}
\label{tab:telegram1}
\begingroup
\setlength{\tabcolsep}{12pt} 
\begin{tabular}{@{}lll@{}}
\toprule
\textbf{Direction} & \textbf{PZD01} & \textbf{PZD02} \\
\midrule
Setpoint (PLC~$\rightarrow$~VFD) & STW1 (control word) & NSOLL\_A (speed setpoint) \\
Feedback (VFD~$\rightarrow$~PLC) & ZSW1 (status word)  & NIST\_A (actual speed) \\
\bottomrule
\end{tabular}
\endgroup
\end{table}

\begin{itemize}
    \item \textbf{STW1} (Control Word~1, 16-bit) drives the inverter state machine. Bit~3 (enable operation), for example, is a field an attacker could rewrite to inject a covert setpoint even when the motor is not enabled; the experiments reported here do not exercise this capability, and forward STW1 unmodified.
    \item \textbf{NSOLL\_A} (Speed Setpoint, 16-bit) carries the reference speed as a normalized fraction of a configured reference, with $4000\mathrm{h}$ (16384) denoting 100\% of the configured reference, here 1800 RPM.%
    
    \item \textbf{ZSW1} (Status Word~1, 16-bit) reports the inverter state (e.g.\ ready, operation enabled, fault active, quick-stop active). It is likewise forwarded unmodified in the experiments reported here.
    \item \textbf{NIST\_A} (Actual Speed, 16-bit) is the measured motor speed, in the same normalization as NSOLL\_A.
\end{itemize}

\begin{figure}[htbp]
    \centering
    \includegraphics[width=0.8\linewidth]{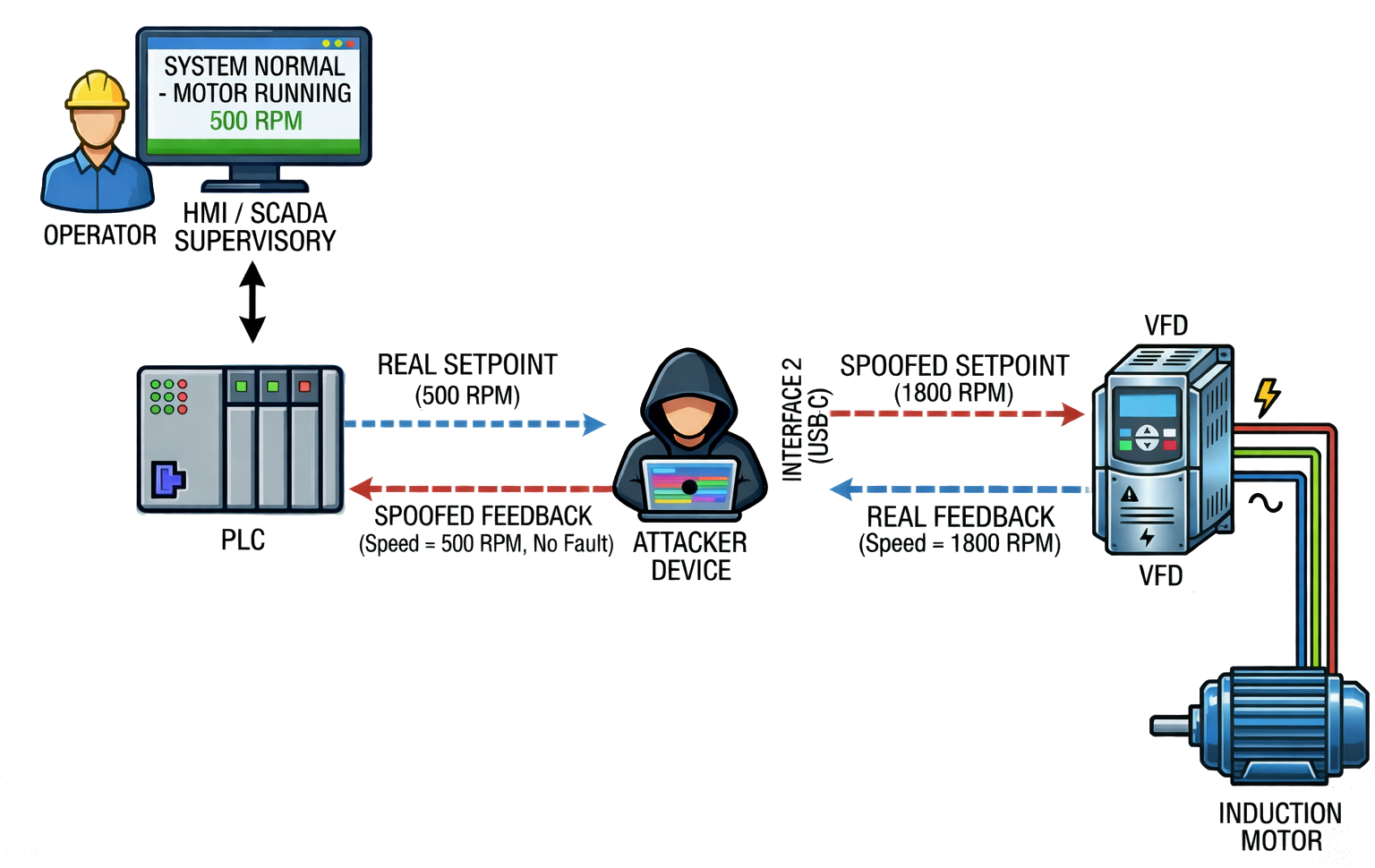}
    \caption{Logical schematic diagram of the Man-in-the-Middle attack. The network switch is omitted to emphasize the logical data flow interception by the attacker between the PLC and the VFD.}
    \label{fig:mitm_schematic}
\end{figure}

\subsection{Man-in-the-Middle Attack}
The attack uses a physical Man-in-the-Middle topology, illustrated in Fig.~\ref{fig:mitm_schematic}. The attacker's station is inserted between the PLC and the network switch that connects the VFD, and the Linux network stack bridges the two Ethernet interfaces, making the interception transparent to the PROFINET protocol.

Deploying this attack requires physical insertion into the PLC--VFD link, but no credentials, engineering software, or knowledge of the drive's configuration. The two physical ports are joined into a single layer-2 bridge with no IP address of their own, so the station is not individually addressable on the segment.

The attack proceeds in three stages:
\begin{enumerate}
    \item \textbf{Interception:} A Python script intercepts PROFINET frames in real time and edits them in place, without disrupting the bridge flow.

    \item \textbf{False Data Injection (FDI):} The script locates the cyclic Telegram~1 frames and rewrites their process-data words in real time. Two capabilities follow from the word layout above:
    \begin{itemize}
        \item \textit{Downlink (setpoint and control).} The attacker overwrites the setpoint to command a speed change.
        \item \textit{Uplink (feedback masking).} To keep the manipulation invisible to the operator, the attacker simultaneously rewrites the feedback word so that the SCADA/HMI observes the response expected for the \emph{unmodified} command. The model-based masking considered here requires an estimate of the plant response.
    \end{itemize}

    \item \textbf{Forwarding:} The modified frames are reinjected. PROFINET cyclic monitoring declares a connection fault after the three consecutive missed cycles configured here.
\end{enumerate}

The uplink masking step is the crux. In the model-based construction studied here, whether the goal is to hide a fault or, as evaluated below, to hide a covert speed change, the quality of the feedback spoof depends on how accurately the attacker models the plant response. The remainder of this paper studies that identification and the residues it leaves behind.

In summary, the attacker can rewrite cyclic traffic in both directions, but the defender evaluated throughout this paper observes only the PLC-side feedback stream and does not jointly validate it against the setpoint; this asymmetry is revisited when the defender's scope is stated explicitly in Section~\ref{sec:artifats}.

\section{Linear Model for Covert System Identification} \label{sec:model}

Unlike theoretical scenarios where the attacker has full access to the plant's states, our attacker relies solely on data exposed over the PROFINET network, so Smith's covert misappropriation~\cite{smith} cannot be applied in its original form.

In our topology, the attacker observes only PROFINET frames; the VFD drives the motor through internal signals that do not appear on the network. The relevant plant is therefore the combined \textit{VFD--motor} system: whereas an induction motor in isolation is conventionally modeled as a first- or second-order LTI system, the VFD's acceleration ramp function dominates the combined system's response to commanded setpoint steps, imposing a rate limit---a nonlinearity that no LTI model can represent.

For the covert setpoint changes considered here, the VFD executes each command at an approximately constant slew rate, i.e., an approximately constant rate of speed change. The measured speed therefore rises almost linearly from its current value toward the new target. Consequently, the time $\tau$ required to execute a step of magnitude $|\Delta v|$ is approximately proportional to $|\Delta v|$.

Because the plant is held at a fixed operating point (1500 RPM), the network carries no excitation, and a passively observed steady-state trace is uninformative: identifiability requires the input to be persistently exciting. The attacker must therefore adopt an \textit{active} identification strategy, injecting its own perturbations and correlating their magnitude with the resulting execution time observed in the feedback. Unlike passive observation, these perturbations are themselves potentially detectable events; the concrete injection protocol is given in Section~\ref{sec:idprotocol}.

The model relates the magnitude of a commanded speed change to the time the drive needs to execute it. Let $v_k$ be the current speed, $v_{k+1}$ the next target, $\Delta v_k = v_{k+1}-v_k$ the signed step, $|\Delta v_k|$ its magnitude, and $\tau_k$ the resulting execution (settling) time. The relationship is modeled as
\begin{equation}
\label{eq:lin}
\tau_k = \alpha\,|\Delta v_k| + \beta,
\end{equation}
with slope $\alpha$ (s/RPM) and intercept $\beta$ (s); $\tau_k$ is in seconds and $\Delta v_k$ in RPM. The timing law is memoryless and cheap to evaluate, hence suitable for real-time covert deployment.

\paragraph{Feedback masks.}
In this paper, a \emph{feedback mask} is the model-based rewriting rule used by the attacker on the feedback word. Let $y(t)$ be the actual VFD feedback under the covert command, and let $\hat{y}_M(t)$ be the response predicted by an attacker model $M$ for the same covert setpoint change. The attacker forwards to the PLC the masked signal
\begin{equation}
    \tilde{y}_M(t)
    =
    y_{\mathrm{nom}}(t) + \bigl(y(t)-\hat{y}_M(t)\bigr),
    \label{eq:mask}
\end{equation}
where $y_{\mathrm{nom}}(t)$ is the feedback that the PLC/operator should observe under the unmodified command. Thus, a perfect mask would make $\tilde{y}_M(t)=y_{\mathrm{nom}}(t)$. Any model mismatch remains visible as the residual
\begin{equation}
    e_M(t)=y(t)-\hat{y}_M(t).
    \label{eq:mask_residual}
\end{equation}
$M$ is called a \emph{static mask} when $\hat{y}_M(t)$ is the constant-ramp-rate prediction \eqref{eq:linear_mask_prediction}, and an \emph{N4SID mask} when $\hat{y}_M(t)$ is generated by a state-space model fitted with N4SID, introduced as a comparison point in Section~\ref{sec:attack_model_comparison}.

\subsection{Identification Protocol}
\label{sec:idprotocol}
To excite the plant while disturbing the operating point as little as possible, the attacker applies a \emph{symmetric bidirectional probing} strategy around the nominal setpoint $v_0 = 1500$~RPM. For a prescribed maximum probing amplitude $s$, each identification round visits a set of targets spread evenly over $[v_0-s,\,v_0+s]$, so the round contains multiple setpoint changes with magnitudes $0<|\Delta v_k|\leq s$; every target is approached from $v_0$ and then returned to it, sampling both acceleration and deceleration at each magnitude. The symmetric pattern keeps the time-averaged setpoint at $v_0$, so the mean process value is undisturbed. The amplitude $s$ is bounded by the stealth constraint: it must stay small enough that the resulting leakage escapes the detectors introduced below. Each perturbation is observed for one settling period, and the attacker fits the affine timing law \eqref{eq:lin} by ordinary least squares over the resulting distinct $(|\Delta v_k|, \tau_k)$ pairs.

\subsection{Reference Parameters from Controlled Steps}

These parameters are a research-side characterisation, not an attacker action: the steps are commanded directly at the drive, over a range far wider than the covert probing of Section~\ref{sec:idprotocol}. The attacker acquires its own model by that protocol, knowing only the four Telegram~1 words of Section~\ref{sec:telegram} --- no drive parameters, motor nameplate data, or ramp settings.

\begin{figure}[htbp]
    \centering
    \includegraphics[width=\linewidth]{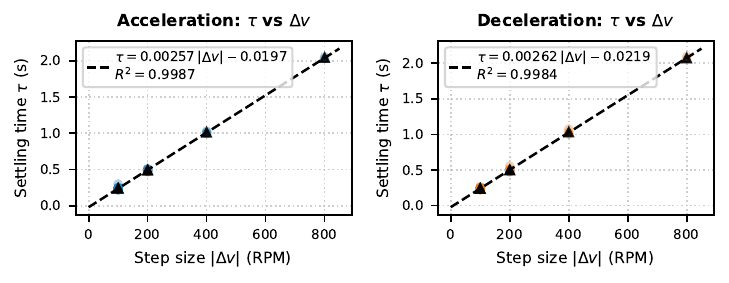}
    \caption{Timing law fitted to the controlled step experiments: settling time $\tau$ vs.\ step size $|\Delta v|$ for acceleration (left) and deceleration (right). Dots: individual measurements; triangles: per-step means; dashed line: fitted model \eqref{eq:model_fit}. The fit tracks the commanded ramp; the terminal overshoot it omits is the residual analyzed in Section~\ref{sec:artifats}.}
    \label{fig:timing_fit}
\end{figure}

Model parameters were estimated from controlled step experiments. Speed steps of $|\Delta v| \in \{100, 200, 400, 800\}$~RPM were applied, and the settling time $\tau_k$ was measured from each setpoint change to the first instant the measured speed reached 99\% of its target. The acceleration set comprises 35 steps (18 at 100~RPM, 11 at 200, 4 at 400, 2 at 800), and deceleration mirrors this with an identical count. Linear regression on the acceleration data yields:
\begin{equation}
    \hat{\tau}_k = 0.00257\,|\Delta v_k| - 0.020~\text{s}, \qquad R^2 = 0.9987.
    \label{eq:model_fit}
\end{equation}
The coefficient $\alpha = 0.00257$~s/RPM corresponds to an effective ramp rate of approximately 389~RPM/s. The small negative intercept ($\beta = -0.020$~s) is not interpreted as a physical delay; accordingly, the ramp predictor introduces no separate dead-time term. Figure~\ref{fig:timing_fit} shows the scatter and fitted lines for both directions, which yield nearly identical parameters.

The fitted linear model defines the static feedback mask used in the experiments. For each commanded step $\Delta v_k$, beginning from the current speed $v_k$ at time $t_k$, the attacker predicts the attack-induced speed trajectory
\begin{equation}
    \hat{y}_{\mathrm{lin}}(t)
    =
    v_k+\operatorname{sign}(\Delta v_k)
    \min\{|\Delta v_k|,(t-t_k)/\alpha\},
    \label{eq:linear_mask_prediction}
\end{equation}
that is, a constant-ramp-rate response clamped at the target, with $\hat{y}_{\mathrm{lin}}$ and $v_k$ in RPM, $t$ and $t_k$ in seconds.\footnote{The clamp makes \eqref{eq:linear_mask_prediction} linear in $t$ only up to the target; \emph{static linear} refers to the affine timing law \eqref{eq:lin} it is built from, not to the trajectory itself.} The intercept $\beta$ does not enter this trajectory; the ramp predictor introduces no separate dead-time term and starts immediately at $t_k$. The attacker then rewrites the feedback word according to the feedback-mask rule in \eqref{eq:mask}, so that only the mismatch between the true VFD response and the predicted response remains visible. Deviations between $y(t)$ and $\hat{y}_{\mathrm{lin}}(t)$, especially the terminal overshoot, constitute the overshoot leakage analyzed in Section~\ref{sec:artifats}.

\section{Detectability at the Control Layer}

\label{sec:artifats}

Pursuing the two-phase objective of Section~\ref{sec:intro}---identify, then attack---the fitted linear model exposes the attacker at two points: during identification, each probe is a potentially observable event; during the covert attack, the model's feedback leaves a residual \eqref{eq:mask_residual} in the feedback channel: the VFD drives the motor to the new target along its configured acceleration ramp, then briefly overshoots before settling, and the static mask reproduces the ramp but not that terminal overshoot. This \emph{overshoot leakage} is the observable a defender monitoring the measured-speed channel can exploit, and it grows with step size. Two detectors are introduced and calibrated, and both phases are evaluated in turn---the identification phase (cascade strategy, Section~\ref{sec:cascade_detection}) and the attack phase (staircase strategy and N4SID comparison).

\subsection{Anomaly Detectors}

Two anomaly detectors are evaluated on the measured-speed residual seen by the PLC after the covert mask is applied.

\textbf{PASAD~\cite{pasad}:} From a sliding window of the last $L$ speed samples, PASAD builds a low-dimensional subspace model of the plant's normal dynamics, then scores each new window by how far it departs from that subspace; this departure score is denoted $D_j$. $L = 500$ samples and subspace dimension $r = 2$ are used. An alarm fires when $D_j$ exceeds the training maximum.

\textbf{CUSUM \cite{Montgomery_2020}:} It accumulates positive and negative drifts $g^+$ and $g^-$ from the training mean, with reference shift $\kappa = 0.5\sigma$, where $\sigma$ is the training standard deviation. An alarm fires when $\max(g^+,g^-)$ exceeds $h$, the maximum accumulated drift observed over the training window.

\subsection{Calibration}
\label{sec:calibration}

Both detectors are trained on a calibration dataset, which covers the motor's start-up and settling to its 1500~RPM operating point. The resulting thresholds are:
\begin{itemize}
    \item \textbf{PASAD:} $\theta_P = 339.7$, training mean $\mu_0 = 1499.9$~RPM.
    \item \textbf{CUSUM:} $h = 317.1$, $\kappa = 0.28$~RPM.
\end{itemize}

\textbf{Threshold calibration. }
The thresholds are set as the maximum departure (PASAD) or drift (CUSUM) observed over the training window, the standard self-calibrating rule for PASAD~\cite{pasad}, applied identically to CUSUM. Importantly, this window includes the motor's own settling transient as it reaches the 1500~RPM operating point from start-up (a $0\to1500$~RPM step response): a brief mechanical overshoot to $\approx$ 1507~RPM above a noise floor of $\sigma = 0.56$~RPM. The excursion is too short to move the training standard deviation appreciably---hence $\kappa = 0.5\sigma = 0.28$~RPM---but it dominates the \emph{accumulated} drift that sets $h$. A detector that alarmed on the plant's normal start-up and re-settling transients would be unusable in practice, so a single threshold spanning all operating regimes \emph{must} sit above the largest benign transient. Because that transient is an upward overshoot, it is $g^+$ that reaches the training maximum $h = 317.1$. The calibration trace holds no benign downward excursion of comparable size---the raw negative-drift maximum is only $5.1$---so $g^-$ stays far below $h$, and the shared threshold is in effect set by the upward direction.

\textbf{Covert window.}
Because that single threshold must sit above the largest benign transient, that same transient defines the attacker's covert window. Overshoot leakage from a small covert step has a similar qualitative form as a benign settling transient, so a perturbation whose leakage stays below the benign envelope cannot be flagged without also alarming on normal start-up and re-settling; under this calibration regime the covert margin is set by the magnitude of the plant's intrinsic transients.

\textbf{Calibration trade-off.}
This conclusion relies on a deployment assumption: the detector is active from cold start and must therefore absorb the start-up transient into its baseline. A deployment that separates the two regimes, either by suppressing monitoring during start-up, or by calibrating start-up and steady operation independently, could apply the steady-state noise floor during normal operation. In our data, excluding the start-up excursion from the training window---the noise floor $\sigma$ itself is unchanged---would yield approximately $\theta_P \approx 2.2$ and $h \approx 5$, shrinking the covert margin by two orders of magnitude; the negative-direction maximum noted above, $h^- = 5.1$, is already at that scale. The cost is a blind interval after every restart, or a mechanism to detect the regime transition.

\subsection{Detectability of the Identification Phase: the Cascade Strategy}
\label{sec:cascade_detection}
Identification is not free: probing the plant requires injecting perturbations, and each perturbation is itself an observable event. A cautious attacker may attempt to bootstrap the model incrementally, using each estimate to mask the next, larger probe. The first unmasked round tests whether zero-knowledge acquisition is feasible; subsequent rounds evaluate masked model refinement. This \emph{cascade} strategy, given in Procedure~\ref{alg:cascade-identification}, is evaluated to determine how far the identification phase can proceed before it becomes detectable.

The cascade protocol executes seven successive identification rounds at probing levels $p_k \in \{0.5, 1, 2, 4, 6, 8, 10\}\%$ of the nominal 1500~RPM, indexed $k = 1,\dots,7$. The first round ($p = 0.5\%$, maximum probing amplitude 7.5~RPM) applies \emph{no} covert mask, as no model has yet been estimated. From $p = 1\%$ onward, the attacker applies the linear ramp mask fitted from the preceding level's VFD response, using the same regression procedure as in Section~\ref{sec:model}.

 \begin{algorithm}[htbp]
\caption{Evaluation of cascade identification under threat monitoring. \Defender{Italic blue} lines are the evaluation, not attacker actions.}
\label{alg:cascade-identification} \footnotesize
\begin{algorithmic}[1] \footnotesize
\Procedure{CascadeIdentification}{$v_0$, $\mathcal{P}$, $\Theta$, $\mathcal{D}$}
    \State Initialize previous mask model $M_{\mathrm{prev}}\gets \emptyset$
    \State Initialize detection log $\mathcal{L}\gets \emptyset$
    \For{each probing level $p\in\mathcal{P}$ in increasing order}
        \State Set maximum probing amplitude $s\gets p\,v_0$
        \State Construct a symmetric bidirectional probing sequence around $v_0$ using multiple step magnitudes $0<|\Delta v_k|\leq s$
        \If{$M_{\mathrm{prev}}=\emptyset$}
            \State Execute the probing sequence without feedback masking
        \Else
            \State Execute the probing sequence while masking feedback with $M_{\mathrm{prev}}$
        \EndIf
        \State \Defender{Record the feedback trace observed by the PLC}
        
        \State \Defender{Compute detector scores using the detector suite $\mathcal{D}$}
       
        \If{\Defender{any detector score exceeds its threshold in $\Theta$}}
            \State \Defender{Mark probing level $p$ as detected}
        \Else
            \State \Defender{Mark probing level $p$ as stealthy} 
        \EndIf   
        \State Observe step increment $\Delta v_k$ and settling times $\tau_k$
        \State Fit a new linear ramp model $M_p$ from pairs $(|\Delta v_k|,\tau_k)$
        \State \Defender{Append $(p,M_p,\text{detector scores})$ to $\mathcal{L}$}
        \State $M_{\mathrm{prev}}\gets M_p$
    \EndFor
    \State \Return fitted models \Defender{ and detection log $\mathcal{L}$}
\EndProcedure 
\end{algorithmic}
\end{algorithm}

The detector scores reported below are not computed on the live-masked trace. Instead, the true VFD response is recorded and the same mask is re-applied offline to reconstruct the trace the PLC would have seen---a step that belongs to this evaluation, not to the attacker.

Offline reconstruction is faithful here because the covert mask rewrites only the feedback word forwarded to the PLC, leaving the VFD's internal speed-control loop untouched. The masked trace is therefore a deterministic function of the recorded VFD response, and reconstructing it offline reproduces the ideal signal-domain feedback the PLC would have received under immediate, lossless rewriting online. This idealization sets aside implementation-level effects such as bridge latency, jitter, or dropped frames; the distinction between offline and online evaluation concerns deployability and computational timing, not the detector scores under this idealization. This point is revisited in Section~\ref{sec:discussion}.

Concretely, for each of five runs the ramp rate fitted at level $p_k$ is extracted and applied as the mask at the next level $p_{k+1}$, and the masked feedback is scored with the PASAD and CUSUM detectors against thresholds $\theta_P = 339.7$ and $h = 317.1$. As in Section~\ref{sec:model}, the mask starts ramping immediately (zero dead-time). Table~\ref{tab:cascade} summarizes results.

The unmasked 0.5\% round is highly detectable: mean PASAD $= 29\,857$ ($88\times\theta_P$). This score dwarfs every later round not because the perturbation is larger---it is the smallest---but because no mask is applied. The detector sees the \emph{entire} symmetric bidirectional probing excursion, a $\pm$7.5~RPM swing about the setpoint. From $p = 1\%$ on, the linear mask subtracts the predicted ramp, and only the post-mask \emph{residual}, the mechanical overshoot the model omits, reaches the detector.

Once a mask is applied, that residual is dominated by mechanical overshoot, which scales with perturbation amplitude. At $p \in \{1, 2\}\%$ (maximum probing amplitudes $\leq$30~RPM) the overshoot is small enough that both detectors remain below threshold in every run (PASAD $\leq 134 < \theta_P = 339.7$): the identification is covert. From $p = 4\%$ onward (maximum probing amplitudes $\geq$ 60~RPM) the overshoot residual exceeds threshold in all runs, and the peak scores have a tendency to grow with probing level ($2485 \to 4387$). The stealth boundary therefore lies between 2\% and 4\% probing level.

These results have two implications. First, the identification phase is most vulnerable at the unmasked bootstrap step: before any model is available, even the smallest probing signal is exposed as a full commanded excursion rather than as a masked residual. This limits our covertness claim to model \emph{refinement}, not zero-knowledge \emph{acquisition}. In a steady-state plant with no ambient excitation, the attacker cannot obtain the initial seed model covertly through the bootstrap probe used here; that seed must come from another source, such as prior knowledge, offline measurements, or passive identification during routine setpoint changes. Second, once a mask is available, detectability is governed by the overshoot boundary, so within this ramp-mask family larger perturbations remain detectable even after model refinement.

 \begin{table}[htbp]
\centering
\caption{Cascade identification detection scores (PASAD $L=500$, $r=2$; CUSUM $\kappa=0.5\sigma$; offline mask, zero dead-time). Values are mean~$\pm$~std over 5 runs; \emph{Det}~=~fraction exceeding any threshold.}
\label{tab:cascade}
\begingroup
\footnotesize
\setlength{\tabcolsep}{9pt} 
\begin{tabular}{@{}ccccc@{}}
\toprule
\textbf{Level} & \textbf{Mask from} & \textbf{PASAD peak} & \textbf{CUSUM peak} & \textbf{Det} \\
\midrule
0.5\%  & ---    & $29\,857 \pm 35$   & $8\,171 \pm 76$   & 5/5 \\
1.0\%  & 0.5\%  & $119 \pm 44$       & $193 \pm 48$      & 0/5 \\
2.0\%  & 1.0\%  & $134 \pm 30$       & $205 \pm 23$      & 0/5 \\
4.0\%  & 2.0\%  & $2\,485 \pm 1\,361$ & $1\,111 \pm 311$  & 5/5 \\
6.0\%  & 4.0\%  & $3\,910 \pm 479$   & $1\,485 \pm 78$   & 5/5 \\
8.0\%  & 6.0\%  & $3\,873 \pm 256$   & $1\,530 \pm 46$   & 5/5 \\
10.0\% & 8.0\%  & $4\,387 \pm 964$   & $1\,603 \pm 164$  & 5/5 \\
\midrule
\multicolumn{2}{@{}l}{\textbf{Threshold}} & 339.7 & 317.1 & --- \\
\bottomrule
\end{tabular}
\endgroup
\end{table}

\subsection{Detectability of the Attack Phase: the Staircase Strategy}

Once a model has been acquired, the attacker moves to the attack phase. The objective is a covert misappropriation of the operating point: driving the motor from its nominal 1500~RPM to 1800~RPM while the PLC-side feedback continues to report steady operation at 1500~RPM. Applying the full 300~RPM change as a single step fails. In this experiment, as in all experiments below, the feedback mask removes the predicted ramp so that only the modeling residual remains visible; even so, the direct step produces a mean peak PASAD score of $1422$ ($4.2\times\theta_P$) and a mean peak CUSUM score of $779$ ($2.5\times h$), and is detected in all 10 trials.

This motivates the \emph{staircase strategy}, given in Procedure~\ref{alg:staircase-attack}: instead of applying one large setpoint change, the attacker decomposes the same total change into smaller increments of size $s$, waiting for the motor to re-settle between consecutive increments. Each increment is masked relative to the current accumulated speed offset.

 \begin{algorithm}[htbp]
\caption{Staircase attack and its evaluation with feedback masking. \Defender{Italic blue} lines are the evaluation, not attacker actions.}
\label{alg:staircase-attack} 
\begin{algorithmic}[1] 
\Procedure{StaircaseAttack}{$v_0$, $v_T$, $s$, $M$, $\Theta$, $\mathcal{D}$}
    \State Initialize commanded speed $v\gets v_0$
    \State Initialize masked feedback trace $\mathcal{Y}\gets \emptyset$
    \While{$v\neq v_T$}
        \State Set next command $v'\gets v+\operatorname{sign}(v_T-v)\min\{s,|v_T-v|\}$
        \State Inject covert setpoint change $v\to v'$
        \State Use mask model $M$ to synthesize the expected ramp response $\hat{y}(t)$
        \State Rewrite the feedback channel using $\hat{y}(t)$ so that only the modeling residual remains observable
        \State Append the resulting masked feedback samples to $\mathcal{Y}$
        \State Wait until the motor re-settles
        \State $v\gets v'$
    \EndWhile
    \State \Defender{Treat $\mathcal{Y}$ as the feedback trace observed by the PLC-side monitor}
    \State \Defender{Compute PASAD and CUSUM scores over $\mathcal{Y}$ using $\mathcal{D}$}
    \If{\Defender{any detector score exceeds its threshold in $\Theta$}}
        \State \Defender{Mark the staircase attack as detected}
    \Else
        \State \Defender{Mark the staircase attack as stealthy}
    \EndIf
    \State \Return \Defender{detector scores and detection outcome}
\EndProcedure
\end{algorithmic}
\end{algorithm}

\begin{figure}[htbp]
    \centering
    \includegraphics[width=\linewidth]{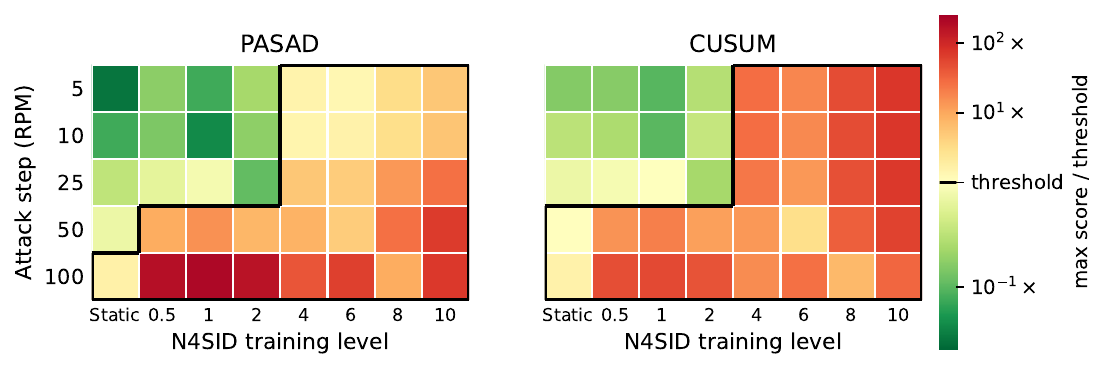}
    \caption{Amplitude-swept N4SID stealth boundary (concealment-selected order~9). Columns: the static mask and N4SID trained at each cascade probing level; rows: covert staircase step size ($n=5$ trials per cell at 5--25~RPM, $n=6$ at 50--100~RPM). Color encodes the worst-case score divided by that detector's own threshold, so the labelled colorbar tick is the detection boundary and both panels share one scale; the black contour marks it within each heatmap. No N4SID training level extends the static mask's boundary, and training at $\geq 4\%$ is detected even at the smallest step.}
    \label{fig:attack_boundary}
\end{figure}

For the static linear mask (\emph{Static} column of Fig.~\ref{fig:attack_boundary}), both PASAD and CUSUM remain below threshold in every trial for $s \leq 25$~RPM. At $s=50$~RPM, the detectors begin to diverge: CUSUM exceeds threshold in 3 of 6 trials, with worst-case score $320>h$, whereas PASAD remains below threshold, with worst-case score $239<\theta_P$. This marks the empirical transition region. At the smallest step size, $s=5$~RPM, the worst PASAD score is only $11$ ($0.03\times\theta_P$), close to the calibration-noise level.

The stealth boundary around $s \approx 25$--50~RPM occurs when the resulting overshoot leakage approaches the plant's benign transient envelope, and the cascade analysis of Section~\ref{sec:cascade_detection} independently locates the same boundary at 2--4\% probing level. A step whose overshoot leakage stays under that envelope is masked by the plant's own dynamics: it is not separated from a routine re-settling event by the evaluated detectors under the adopted calibration. A defender could only compress this window by replacing self-calibrated thresholding with a model that distinguishes attack-induced from commanded transients. An attacker who has characterized the plant's transient envelope can predict the maximum covert step size before initiating the attack.

\subsection{Attack-Phase Model Comparison}
\label{sec:attack_model_comparison}

To test whether this boundary is an artifact of the static model being too simple, the static mask is replaced with N4SID state-space masks and the staircase evaluation is repeated. 

N4SID (Numerical algorithms for Subspace State Space System Identification)~\cite{vanoverschee1994n4sid} estimates a discrete-time state-space model, $x_{k+1}=Ax_k+Bu_k$ and $y_k=Cx_k+Du_k$, directly from input--output data; unlike the ramp predictor it carries state across cycles and can therefore represent transient dynamics the static map omits---but only linear ones. Two choices govern the comparison, the training amplitude and the model order $n$ (the state dimension), and both are swept. A mask is fitted at each cascade probing level $p \in \{0.5,1,2,4,6,8,10\}\%$ and applied to every staircase run. The training-amplitude sweep ranges from 7.5 to 150~RPM, spanning amplitudes below and above most of the evaluated attack steps.

\begin{figure}[tbp]
    \centering
    \includegraphics[width=0.9\linewidth]{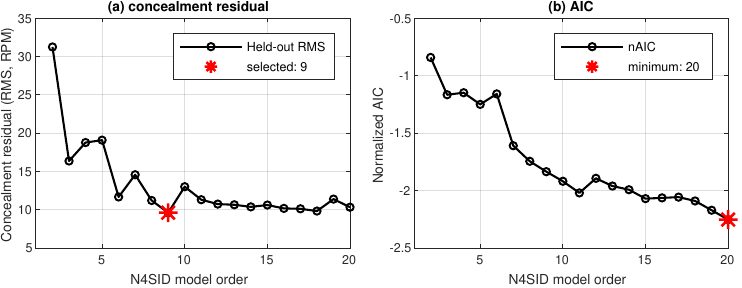}
    \caption{N4SID order selection on the largest-amplitude (10\%) training data. \textbf{(a)} The held-out RMS concealment residual, used as a detector-agnostic proxy for concealment, has an interior minimum at order 9. \textbf{(b)} AIC, computed on the prediction error, falls to the sweep boundary (order~20): it optimizes predictive fidelity, not concealment. The attacker selects for~(a).}
    \label{fig:matched_order}
\end{figure}

The order is selected for the attacker's actual objective, concealment, not identification. The conventional criterion, the Akaike information criterion (AIC)~\cite{akaike1974newlook}, scores a model by its prediction error penalized by the number of free parameters; at the sample counts available here it decreases to the sweep boundary (order~20, Fig.~\ref{fig:matched_order}b): it rewards a predictive fidelity the attacker does not need. The \emph{held-out concealment residual} is instead minimized---the RMS departure, on a validation split, of the masked feedback from the calibration baseline---used here as a detector-agnostic proxy for concealment. This criterion has a minimum at order~9, which is used for every training level.

\begin{figure}[tbp]
    \centering
    \includegraphics[width=0.9\linewidth]{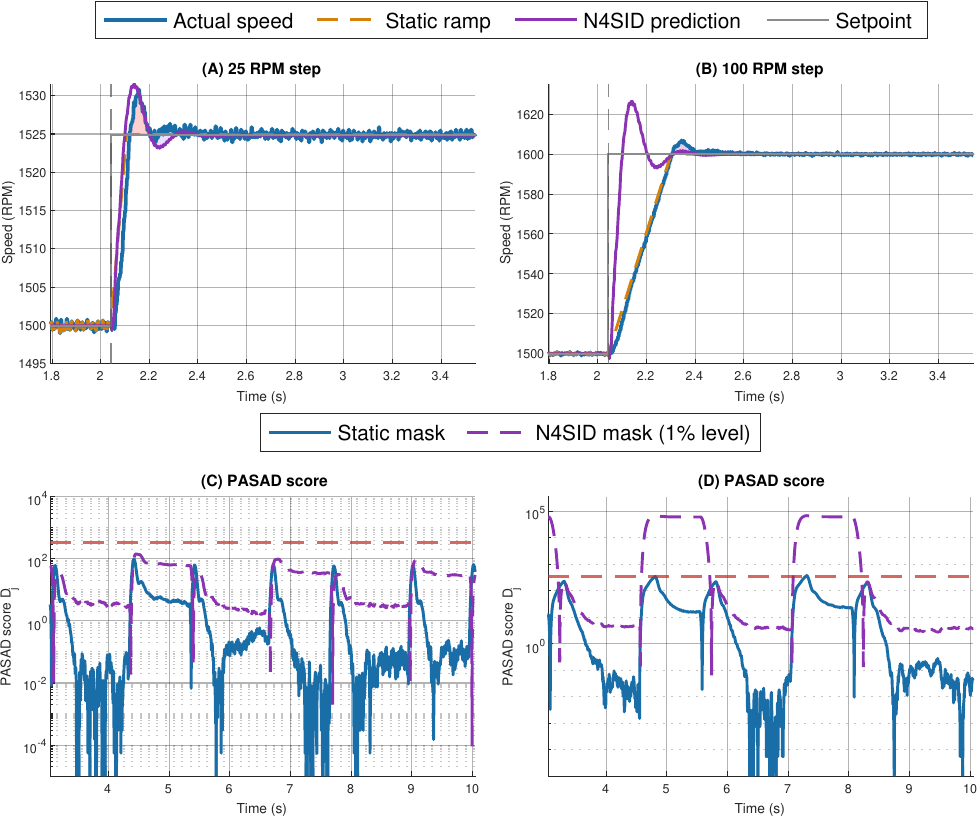}
    \caption{Overshoot leakage: static vs.\ N4SID mask (order~9, 1\% cascade training).
    \textbf{(A)--(B)} Actual VFD speed (blue), static ramp prediction (orange, dashed), N4SID prediction (purple), and setpoint for a 25~RPM and 100~RPM covert step.
    \textbf{(C)--(D)} PASAD score $D_j$ for the static mask (blue, solid) and N4SID mask (purple, dashed); red dashed line: threshold $\theta_P = 339.7$.}
    \label{fig:overshoot_leakage}
\end{figure}

Figure~\ref{fig:overshoot_leakage} shows the mechanism on single traces, for the static mask and the N4SID mask trained at the 1\% level. At a 25~RPM covert step both stay below threshold ($D_j = 99$ static, $D_j = 215$ N4SID). At 100~RPM both are detected, but the N4SID score ($D_j = 71\,303$) is $189\times$ the static score ($D_j = 377$): the 1\%-trained model extrapolates poorly to a step $\approx$6.7$\times$ larger than its training excitation ($\pm$15~RPM probe). Fitting a more complex model at the training amplitude therefore does not imply better concealment under larger covert steps.

Even so, none of the evaluated N4SID masks extends the static mask's stealth region (Fig.~\ref{fig:attack_boundary}). The static map, built from the reference timing law of Section~\ref{sec:model}, is never outperformed: on PASAD it alone keeps the 50~RPM step below threshold, and on CUSUM it crosses only marginally there ($320$ against $h = 317.1$, in 3 of 6 trials). Masks trained at small amplitudes (0.5--2\%) \emph{match} the static map up to 25~RPM, where every mask stays stealthy, but lapse at 50~RPM where the static map still holds. Masks trained at 4\% probing level and above are detected at \emph{every} step size, including the smallest 5~RPM covert step: a 10\%-trained mask leaks a worst-case CUSUM score of $2.3\times10^{4}$ ($74\times h$) for a 5~RPM step, whereas the largest staircase step in the amplitude sweep, 100~RPM, only just crosses ($\approx 1.5\times$). Among the evaluated state-free ramp and LTI subspace models, higher complexity does not push the boundary outward.

\medskip
\noindent\textbf{Remark 1 (the control-layer boundary is structural).} The cascade analysis (Section~\ref{sec:cascade_detection}), the staircase attack, and the amplitude-swept N4SID comparison exhibit a consistent transition around the same transient envelope: leakage below it remains undetected, whereas crossing it increasingly drives the detector scores beyond threshold. This boundary is structural under the self-calibrated, cold-start monitoring assumed here, which must absorb the motor's start-up overshoot into its baseline. A defender that instead gated monitoring around restarts ($\theta_P \approx 2.2$, Section~\ref{sec:calibration}) would compress it sharply, so the margin is a property of the plant \emph{and} of the detector's calibration regime. The leak is not second-order ringing but the terminal overshoot of the drive's configured acceleration ramp (Section~\ref{sec:model}): in Fig.~\ref{fig:overshoot_leakage}B the static map cancels the near-linear rise and leaks only the overshoot, whereas the N4SID prediction, which cannot represent the rate-limit nonlinearity, departs from the actual response across the whole transient. 
\section{Discussion} \label{sec:discussion}

The results expose a narrower but stronger trade-off than the usual intuition that a more accurate model should be stealthier but computationally heavier. At the control layer the attacker is exposed by what the model omits, and Remark~1 shows that added complexity relocates that exposure rather than removing it.

Zeroing the residual would require a model that reproduces the rate-limited rise and its overshoot, which the evaluated masks cannot supply. A natural objection is that a richer model might be difficult to evaluate within the 2~ms PROFINET cycle, which would place the binding trade-off in per-cycle compute rather than in detectability.

\begin{figure}[tbp]
    \centering
    \includegraphics[width=\linewidth]{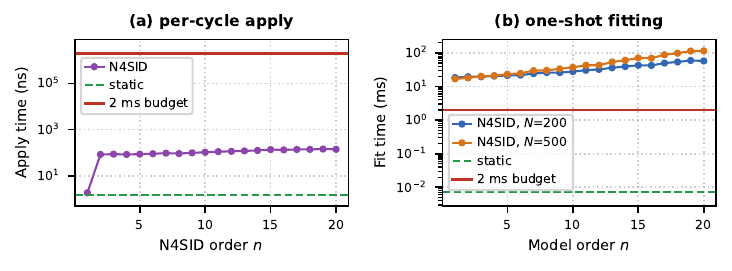}
    \caption{Per-cycle apply cost (left) and one-shot model-fitting cost (right) for the static linear mask and N4SID at orders 1--20 (mean~$\pm1\sigma$, 10 repetitions, log scale). Apply cost is negligible for both at every order; the asymmetry lies in fitting, where N4SID is equivalent in duration to 8--59 PROFINET cycle periods across the full order sweep.}
    \label{fig:realtime_cost}
\end{figure}

\medskip
\noindent\textbf{Remark 2 (per-cycle cost is not the discriminator).} We time only the arithmetic kernel each mask evaluates once per cycle: the static multiply-and-clamp, or one state-space propagation step. The surrounding bridge path (packet interception, parsing, reinjection) is common to both masks and cannot separate them. Figure~\ref{fig:realtime_cost} reports both costs over the full order sweep, measured in MATLAB on the attacker station of Section~\ref{sec:arch}: the static mask applies in 1.4~ns and the heaviest measured N4SID propagation (order~19) in 141~ns, both under 0.01\% of the 2~ms PROFINET budget. The asymmetry lies instead in \emph{fitting}: static OLS on 35 pairs takes $\approx 0.005$~ms ($<0.01$ cycles), whereas N4SID on 200--500 samples takes 16--118~ms across orders~1--20, equivalent in duration to 8--59 PROFINET cycle periods. What distinguishes the two models is the cost of building them, not of running them.

\medskip
If the cost that matters is fitting, what matters next is how often the attacker must pay it. Our experiments treat the plant as time-invariant, and the fitted timing law reproduces across disjoint sets of runs (Section~\ref{sec:limitations}), so within our horizon it is paid once. A plant whose parameters drift would be a different regime; the relevant drift there would be \emph{parameter} drift at a fixed setpoint (thermal effects, mechanical wear, load variation), not setpoint drift, which would supply passive excitation and reduce the need for active re-identification.

\medskip
\noindent\textbf{Remark 3 (re-identification under drift).} Drift was not observed in our experiments, and the fitted timing law reproduces across sessions, so within this horizon a mask can be identified once and reused. Should the plant's parameters drift, however, even the static model---the stronger covert choice by Remark~1---would cease to be a one-time cost. Tracking drift requires new identification pairs, so the attacker must re-probe, and each re-probe would meet the same control-layer stealth boundary measured in the cascade and staircase experiments: sufficiently small perturbations stay hidden, larger ones leak the overshoot residual. Building the model offline and deploying only when it is ready avoids the exposure of re-probing, but not the drift of the model away from the plant: by \eqref{eq:mask_residual} the residual \emph{is} the model error, so a stale mask can increase the leakage.

\subsection{Scope and Limitations}
\label{sec:limitations}

The assumptions behind the results above are stated where they are used; they are collected here.

\textbf{Plant and operating point.} All experiments use a single VFD--motor pair held at a single 1500~RPM operating point. The timing-law fit, detector calibration, cascade, and staircase each draw on a disjoint set of runs, so the reported numbers do not depend on any single session; that independence, however, does not by itself extend them beyond this testbed.

\textbf{Defender model.} The monitored quantity is the PLC-side feedback stream alone. The thresholds follow the self-calibrating, cold-start rule of Section~\ref{sec:calibration}; as quantified there, a deployment that gated monitoring around restarts would shrink the covert margin by roughly two orders of magnitude. The boundary is therefore a property of the plant \emph{and} of this calibration regime, not of the plant alone.

\textbf{Evaluation procedure.} Masks are reconstructed offline under the idealization of immediate rewriting (Section~\ref{sec:cascade_detection}), which sets aside bridge-level latency, jitter, and frame loss. The covertness claim covers model \emph{refinement} rather than zero-knowledge \emph{acquisition}: the unmasked bootstrap probe is detected, so the seed model must come from elsewhere. The attack-phase masks are built from the reference timing law of Section~\ref{sec:model}, which is fitted from steps well above the stealth boundary and is therefore more accurate than the covert probing of Section~\ref{sec:cascade_detection} yields. The attack-phase results accordingly assume an already-acquired model; they are a favourable case for the attacker, not a characterisation of end-to-end covert operation.

\textbf{Model class.} We evaluate model-based feedback masking; covert strategies that do not rely on a plant model are out of scope. Two mask families are evaluated: the state-free ramp predictor, whose clamp is a fixed saturation rather than a fitted nonlinearity, and LTI subspace models fitted by N4SID. Fitted nonlinear and learning-based masks are untested: the order sweep indicates that increasing order alone does not remove the boundary within the evaluated LTI class, but a mask matched to the rate-limited transient \emph{and} its terminal overshoot remains future work.


\section{Conclusion} \label{sec:conclusion}

In this paper, we studied covert system identification under timing constraints. A simple static linear model approximates the VFD--motor response accurately through the commanded ramp ($R^2 \approx 0.999$) at negligible computational cost, making it attractive for real-time MitM deployment. Its limitation is structural: because the model omits the plant's terminal overshoot, each covert step leaks a residual that becomes detectable once the resulting leakage exceeds the benign settling envelope admitted by the detector calibration. The cascade and staircase experiments locate this boundary consistently, and the amplitude-swept N4SID comparison shows that a more complex linear model does not remove it.

The resulting trade-off is not primarily about per-cycle computation, which is negligible for both static and N4SID masks, but about identification cost and robustness across amplitudes: the static model costs orders of magnitude less to fit and was never outperformed by a richer mask in our experiments. These results are bounded by the scope stated in Section~\ref{sec:limitations}. The natural next step is to look outside the LTI class: a nonlinear mask able to reproduce the rate-limited rise together with its terminal overshoot would remove the residual this paper measures, and the open question is whether such a mask can be acquired within the covert data budget the stealth boundary allows.

Future work should test the non-stationary, drifting-plant setting of Remark~3 directly: measuring how parameter drift affects the residual, characterizing the stealth boundary under different detector calibrations, formalizing the staircase as a detection-constrained optimization.

\section*{Acknowledgments}

This research was partially funded by Clavis FINEP/PlatCiber, under FINEP reference 1488/22. Additional support was provided by CAPES (Finance Code 001); FAPERJ under grants E-26/204.268/2024 and E-26/260.168/2026; CNPq under grants 424622/2021-1, 403601/2023-1, 315106/2023-9, 444956/2024-7, and BPP/PQ-Sr 313335/2022-2; and FCT through the LASIGE Research Unit, ref.\ UID/00408/2025, DOI: \url{https://doi.org/10.54499/UID/00408/2025}.

\bibliographystyle{splncs04}
\bibliography{references}

\end{document}